\documentstyle[preprint,aps]{revtex}

\begin{document}

\title
{Study of the nucleon-induced preequilibrium reactions
in terms of the Quantum Molecular Dynamics}

\author{
Satoshi Chiba$^{(1)}$,
Mark B. Chadwick$^{(1,2)}$,
Koji Niita$^{(1,3)}$,\\
Toshiki Maruyama$^{(1)}$,
Tomoyuki Maruyama$^{(1)}$ and
Akira Iwamoto$^{(1)}$}

\address{
$^{1}$Advanced Science Research Center,
Japan Atomic Energy Research Institute, \\
Tokai-mura, Naka-gun, Ibaraki-ken 319-11, Japan\\
$^{2}$Nuclear Data Group, University of California,
Lawrence Livermore National Laboratory, \\
Livermore, California 94550, USA\\
$^{3}$Research Organization for Information Science \& Technology \\
Tokai-mura, Naka-gun, Ibaraki-ken 319-11, Japan}
\maketitle
\begin{abstract}
The preequilibrium (nucleon-in, nucleon-out) angular distributions of
$^{27}$Al, $^{58}$Ni and $^{90}$Zr have been analyzed in the energy
region from 90 to 200 MeV in terms of the Quantum Moleculear Dynamics
(QMD) theory.  First, we show
that the present approach can reproduce the measured (p,xp') and
(p,xn) angular
distributions leading to continuous final states without adjusing any
parameters.
Second, we show the results of the detailed study of the preequilibrium
reaction processes;
the step-wise contribution to the angular distribution, comparison
with the
quantum-mechanical Feshbach-Kerman-Koonin theory, the effects of
momentum
distribution and surface refraction/reflection to the quasifree
scattering.
Finally, the present method was used to assess the
importance of
multiple preequilibrium particle emission as a function of
projectile energy up to 1 GeV.

\end{abstract}



\section{Introduction}\label{introduction}

The nucleon-induced nuclear reactions leading to continuum states
at intermediate (E$_{inc} \ge $100MeV)
energy range are characterized by a reaction mechanism known as the
preequilibrium process\cite{gadioli}.  In this process, particle
emissions take place
from simple particle-hole configurations populated as a result of a
sequence of nucleon-nucleon interactions before the statistical
equilibrium
is attained.  The angular distribution of the particles
emitted from this process has generally a smooth forward
peaking which
is intermediate in character between the direct and compound nuclear
processes.
As the energy of the projectile increases, the number of particles
emitted from the preequilibrium mechanism is increased and exceeds
one (which is therefore called as the
multiple preeqilibrium emission process, MPE process).
At very high energy, the reaction
is often referred to as the "spallation" reaction, in which the
average
multiplicity of ejectile exceeds several or larger.

Study of the preeqilibrium nuclear reactions has been an active field
since the pioneering work of Goldberger\cite{goldberger} and
Metropolis\cite{metroplis} based on the cascade model, and of
Griffin\cite{griffin} based on the exciton
model.  Various refinments on
these approaches as
well as new models both of semi-classical and quantum-mechanical
followed (see for example Ref. \cite{gadioli}).
The semi-classcial models have been applied to analyze the
energy spectra of preequilibrium particles on the outset.  Later
they have been improved to take account of the
angular distributions of the preequilibrium process; the
exciton model was improved to the generalized
excition model\cite{mant,akk1,iwahara} and the geometry dependent
hybrid
model\cite{blann84}.
The cascade model has been able to calculate the angular distributions
 based
on the Monte-Carlo technique.  Furthermore, a semi-classical distorted
 wave
theory was proposed by Luo and Kawai\cite{luo,watakawa} who have
combined the concept
of quantum distorted-wave and the cascade model.  They have applied
this theory to calculate the
1-step double-differential cross sections.  Extension to the 2-step
process was also formulated\cite{kawai}.
Althouth these theories gave overall agreements with the data, there
are still
open questions which need further investigation for a better
understanding of the preequilibrium reaction processes.
For example,
the backward angular distributions calculated by the semi-classical
theories
are often
considerably smaller than the measured values.  Various conjectures
have been made to account
for this
problem\cite{blann84,watakawa,costa,chadwick91,haneishi,chadwick94};
the refraction effects at the nuclear surface,
quantum diffraction,
high momentum component in the momentum distribution, multi-step
effects, MPE, etc.  So far, no simple answer seems to resolve this
problem.
The same problem
persists even in the quantum-mechanical Feshbach-Kerman-Koonin (FKK)
theory\cite{feshbach};
one and the only quantum-mechanical preequilibrium theory which is
able to calculate
the multi-step direct process up to any number of steps at present.
In the FKK theory, furthermore,
there are some other open problems, e.g., the transition between the
unbound
and bound states (P $\iff$ Q transition) as studied recently by
Watanabe et al.\cite{watanabe95}, and use of the normal and non-normal
 DWBA matrix elements
in the calculation of multistep direct components\cite{udagawa}.
On the contrary, the cascade model has a problem at both the very
forward
and backward angles,
where the calculated values are noticeably smaller than the
experimental
data.
Moreover, the number of
particles emitted from the preequilibrium process is limited to only
1 or 2 in many preequilibrium theories proposed so
far\cite{chadwick94,blann83}; an assumption which is questionable when
 the projectile
energy becomes higher and higher.

The purpose of this paper is to study the angular
distributions and  MPE process in the preequilibrium (nucleon-in,
nucleon-out)
 reaction in terms
of a reaction theory based on the molecular dynamics technique, the
Quanum Moleculear Dynamics (QMD)\cite{aichelin,maru,peilert}.  The QMD
 theory includes, in a self-consistent
way, many important aspects in understanding the nucleon-induced
reaction mechanisms at intermediate energy range, i.e.,
1) the realistic momentum distribution of nucleons inside nuclei
(including high-momentum component),
2) entrance/exit channel refraction,
3) Coulomb deflection,
4) multistep process,
5) MPE process,
6) variation of the mean-field potential due to particle-hole
excitation and particle emission,
7) transition between unbound and bound states (P $\iff$ Q transition
in
FKK theory), and
8) energy-dependent, anisotropic N-N elastic and inelastic scattering
including the Pauli-bocking effect.
These features make QMD a useful tool to study
the
nucleon-induced preequilibrium processes in a systematic manner
as was first demonstrated by Peilert et al.\cite{peilert}.  We are
willing to show calculations of better statistics to check its
ability at very backward angles for energetic ejectiles which was
not clear in Ref.\cite{peilert}.  Furthermore, we will clarify the roles
of the step-wise contributions, MPE contributions, the momentum
distribution and surface refraction/reflection to understand
the basic physics of the preequilibrium reactions.

In this paper, we use the method as formulated in Ref. \cite{niita},
extend the analyses given in Ref.
\cite{chadchiba}, and will show that
the present QMD approach gives a consistent description of the
preequilibrium
(p,xp')
and (p,xn) angular distributions of $^{27}$Al, $^{58}$Ni and $^{90}$Zr
 targets
at 90 to
 200 MeV energy range in the
entire angular region without any fitting procedure.  Based on the
excellent
agreement with the data, we then proceed to study some of the open
problems
left in the preequilibrium processes.
In section \ref{brief},
we will give a brief explanation of the QMD to show the essential
feature of
our model.  In section
\ref{results}, we compare our results with experimental data
and
predictions of the FKK theory to find similarities and differences
between
these two theories.  We will then give
a further discussion on the angular distribution of the quasifree
scattering (QFS), and the energy dependence of the MPE process.

\section{Brief explanation of The Quantum Molecular Dynamics}
\label{brief}

\subsection{Equation of motion}
We start from representing each nucleon (denoted by a
subscript {\it i}) by a Gaussian wave packet
in both the coordinate and momentum spaces in the following way:

\begin{equation}
f_i({\bf r}, {\bf p}) = 8 \cdot \exp \left[ -\frac{({\bf r}-{\bf
R}_i)^2}{4L}
-\frac{2L(\bf{p}-\bf{P}_i)^2}{ \hbar ^2} \right]
\label{eq:phase-space}
\end{equation}
where ${\it L}$ is a parameter which represents the spacial spread of
a
wave paccket,
${\bf R}_{\it i}$  and ${\bf P}_i$ corresponding to
the centers of a wave packet in the coordinate and momentum spaces,
respectively.  The total
one-body phase-space distribution function is taken to be simply a sum
 of these single-particle wave packets.  The equation of
motion of ${\bf R}_i$  and
${\bf P}_i$ is given, on the basis of the time-dependent variational
principle, by the
Newtonian equation:

\begin{equation}
\dot{\bf R}_i=\frac{\partial H}{\partial {\bf P}_i}, ~~~~~
\dot{\bf P}_i=-\frac{\partial H}{\partial {\bf R}_i},
\label{eq:eos}
\end{equation}
and the stochastic N-N collision term as will be described below.  We
have adopted the Hamiltonian
$\it{H}$ to
consist of the relativistic kinetic+mass energy and the Skyrme-type
effective N-N
interaction\cite{skyrme} plus Coulomb and symmetry energy terms:

\begin{eqnarray}
H~=~& &\sum_{i}\sqrt{m^2_i~+~\bf{P}^2_i} \nonumber \\
& &{} +\frac{1}{2} \frac{A}{\rho_0} \sum_{i}<\rho_i>
+\frac{1}{1+\tau} \frac{B}{\rho_0^\tau} \sum_{i}<\rho_i>^\tau
\nonumber \\
& &{} +\frac{1}{2}\sum_{i,j(\neq i)} c_i c_j \frac{e^2}{\mid
\bf{R}_i-\bf{R}_j \mid}
   {\rm erf} \left( \mid \bf{R}_i-\bf{R}_j \mid /\sqrt{4 {\it L}}
\right) \nonumber \\
& &{} +\frac{C_s}{2 \rho_0} \sum_{i,j(\neq i)} \left( 1-2 \mid c_i-c_j
 \mid \right) \rho_{ij},
\label{eq:hamiltonian}
\end{eqnarray}
where "erf" denotes the error function, and the $c_i$ is 1 for proton
and 0 for neutron.
The other symbols in eq.(\ref{eq:hamiltonian}) are defined as:

\begin{eqnarray}
\rho_i({\bf r})&\equiv &\int \frac{d{\bf p}}{(2 \pi \hbar)^3}f_i({\bf
r},{\bf p})\nonumber \\
{}&=& (2 \pi L)^{-3/2} \exp \left[-({\bf r}-{\bf R}_i)^2/2L \right]
\label{eq:rhoi}
\end{eqnarray}
and

\begin{eqnarray}
<\rho_i>&\equiv&\sum_{j (\ne i)} \rho_{ij} \equiv \sum_{j (\ne i)}
  \int d {\bf r} \rho_i({\bf r}) \cdot \rho_j({\bf r})\nonumber \\
 {}&=& \sum_{j (\ne i)}(4 \pi L)^{-3/2} \exp \left[-({\bf R}_i-{\bf
R}_j)^2/4L \right].
\label{eq:hoho}
\end{eqnarray}
The symmetry energy coefficient $C_s$ is taken to be 25 MeV.  The four
 remaining parameters,
the saturation density  $\rho_0$, Skyrme parameters $A$, $B$ and
$\tau$
are chosen to be 0.168 fm$^{-3}$, $-124$ MeV, 70.5 MeV and  4/3,
respectively.
These values give the binding energy/nucleon of 16 MeV at the
saturation density $\rho_0$ and the compressibility of 237.7 MeV (soft
 equation of state, EOS) for nuclear matter limit.
The only arbitrary parameter in QMD, i.e., the width parameter {\it
L}, is fixed to be 2
fm$^2$ to give stable ground state of taget nuclei in a wide mass
range.  These
values are taken from our previous paper\cite{niita}.

\subsection{The collision term and the Pauli blocking}
     The stochastic nucleon-nucleon collision is taken into
consideration as similar to the cascade
model\cite{niita}:  When the impact parameter of two nucleons is
smaller than
a value of $ \sqrt{{\sigma}/{\pi}}$ where $\sigma$ denotes
the energy-dependent N-N cross section, an elastic or inelastic N-N
collision
takes place.
We adopt a parametrization of N-N cross sections\cite{niita}
which is similar to that of Cugnon\cite{cugnon} to
take account of the
in-medium effects which reduces the absolute magnitude and
forward-peaking of
the N-N cross sections.
The angular distribution of the elastic scattering was selected by the
Monte-Carlo sampling method.

The Pauli blocking of the final phase-space is checked after each
collision.
The blocking probability is calculated in the same way as
the collision term
in the Vlasov-Uehling-Uhlenbeck theory\cite{bertsh}.

The parameters in the N-N cross sections
were fixed in Ref.\cite{niita} and are
used in this paper.  Together with the parameters of the one-body
dynamics given in the previous subsection, the equation-of-motion
of the QMD is uniquely determined.

\subsection{The ground state}
     The ground state of the target nucleus is generated by packing
${\bf R}_i$
and ${\bf P}_i$ randomly based on the
Woods-Saxon type distribution in the coordinate space and
corresponding local Thomas-Fermi
approximation in the momentum space, seeking a configuration to
reproduce the
binding energy
calculated by the liquid-drop model within a certain ($\pm$ 0.5MeV)
uncertainty.

The average distribution of the ${\bf P}_i$
obtained for $^{90}$Zr is shown in Fig.~1 as the solid histogram.
Experimental nucleon momentum distribution in nuclei is
parametrized by a superposition of
2-Gaussians\cite{haneishi} as

\begin{equation}
\rho(p)~=~N_{1} \left( e^{(-p^2/p_{0}^2)}~+~\epsilon_{0}
e^{(-p^2/q_{0}^2)} \right)
\label{eq:haneishi}
\end{equation}
where
$N_{1}$ is just a normalization constant, and
the parameters $p_{0}$ and $q_{0}$ are related to the Fermi momentum
$p_{F}$ via

\begin{eqnarray}
p_{0}~&=&~\sqrt{2/5} p_{F},~~{\rm and}\nonumber \\
q_{0}~&=&~\sqrt{3} p_{0}.
\label{eq:p0q1}
\end{eqnarray}
This distribution is plotted as a broken curve in Fig. 1, where the
parameter
$\epsilon_{0}$ has been taken to be 0.07; about the mid-point
of the range of this parameter recommended by Haneishi and Fujita.
The
nucleon momentum distribution in the QMD calculation has a similar
shape to this
2-Gaussian distribution, while the commonly adopted uniform Fermi gas
distribution is
just a simple square-shaped function which vanishes above the Fermi
momentum.
The most significant difference among these distribution is the
presence
of the high-momentum component in the former two distributions which
is not
present in the uniform Fermi gas model.
The presence of the high-momentum component is a common feature of
finite-nucleon systems. As a matter of fact, the momentum distribution
in QMD has a very similar shape to the one obtained by the Hartree-Fock
theory as compared in Fig. 6(b) of Ref. \cite{niita}.
It is well
known that the high momentum component enhances the backward angular
distributions\cite{haneishi}, and
as will be shown later, we obtain the same conclusion from our QMD and
1-step Monte-Carlo simulations.  However, the effect of the difference
 in the momentum distribution
on the angular distribution of the primary particles emitted from the
quasi-free scattering process was not very remarkable
except at the very forward and backward
angles, as will be
discussed in the next section.
It may worth noting here that the ground state in QMD as obtained
in our work remains stable even with the high-momentum tail.

\subsection{Decomposition into step-wise contribution in multistep
reactions}
     For later discussion of the multistep reaction, it will be
convenient to give a
definition of step number in the QMD calculation which should reflect
the number of
 collisions responsible for emission
of a nucleon.  First we assign a step number of 0 to each nucleon in
the target nucleus.
After a nucleon collides with incident nucleon, we set collision
number 1 to both
nucleons, inhibitting a collision between nucleons of collision number
 zero pair.
Also, we prohibit successive collisions by the same partner.
The rule of the change of the step number for
each nucleon is that, if two nucleons
{\it i} and {\it j} having step numbers $s_i$ and $s_j$ make a
collision, the step numbers
of both particles are modified to be $s_i + s_j + 1$.
We then identify that a nucleon is emitted from the {\it n}-step
process if an isolated nucleon
emitted from the nucleus has a step number of {\it n}.

     As explained above, the first collision takes place only between
the projectile and a nucleon in the
target nucleus as expected intuitively.  If one or two of these
nucleons are emitted without experiencing
further collisions, they contribute to the 1-step process.  If, on the
 other hand, either
of these nucleons makes a further collision in the nucleus, and one or
 two of these nucleons
involved in the second collision are emitted without
further collision, they are classified as the 2-step
process.
In the FKK theory, on the contrary, the 1-step direct cross
section is calculated
by means of
the normal DWBA method averaged over many final 1p-1h states, that
is caused as a result of having 1 collision between the projectile
and a nucleon in the target.  The m-step FKK direct component is
calculated by a folding integral of the (m-1)-step and 1-step
cross sections, that results after a nucleon under interest
has experienced the m-th collision in the system.
The difinitions of the step number in QMD and FKK coincide
up to the step number of 3.  Beyond the 3-step process,
however,
those definitions become slightly inconsistent because
QMD
includes collisions between collided nucleons which are not
present in FKK approach, although the probability of having
such collisions in QMD is not very large.

\subsection{Calculation of the Cross Section}
\label{cross section}

     In the calculation, many events having different impact
parameter
were generated.  The impact parameter has been selected from a
uniform distribution between 0 and a maximum value which was taken to
be slightly
bigger than the nuclear radius.
The energy and direction of motion are stored event
by event for every nucleon that becomes free (isolated from other
nucleon),
and finally the double-differential
cross section was calculated as

\begin{equation}
\frac{\partial^{2}\sigma}{\partial E \cdot \partial \Omega}~=~\int
2 \pi b \cdot
 \left< M(E,~\Omega,~b) \right> db
\label{eq:cross section}
\end{equation}
where
$\left< M(E,~\Omega,~b) \right>$ denotes the average multiplicity of
the particle under
interest (neutron or proton) emitted
in the unit energy-angular
interval around
$E$ and $\Omega$ for the impact paramter $b$ event.

Typically, 50000 events were generated to get a reasonable statistics
in the
step-wise double-differential cross section.  In the calculation, the
parameter
has been fixed to the same values as in Ref. \cite{niita},
without any adjustment.

\section{Results and Discussion}
\label{results}

\subsection{Comparison with experimental data}

     The calculated double-differential $^{58}$Ni(p,xp') cross
sections
for incident energies at 120 and 200
MeV, and the $^{90}$Zr(p,xp') and (p,xn) cross sections at 160 MeV are
compared in Figs. 2 and 3 with experimental
data\cite{fortsch,richter,scobel}.  The data have been shifted by the
amount denoted
in the parentheses.  Agreement of the present calculation with the
measured values is quite satisfactory from the very forward to
backward angles,
showing a basic ability and usefulness of our QMD approach to
investigate the N-A
reaction mechanisms in this
energy regime.  The problem of the underestimation at the backward
angles
in the semi-classical models\cite{blann84,costa} and the problems
in the cascade model\cite{yoshizawa,takada,bertini}
at the very forward and
backward
angles are not present in the QMD approach.
 It must be also noticed that the QMD theory reproduces both the
(p,xp') and (p,xn) cross sections simultaneously with a single set of
parameters.
  This is a clear advantage of this approach
over, e.g., the multistep direct FKK theory in which strength of the
effective N-N cross
section (the V$_0$ parameter) must be adjusted depending on the
projectile,
ejectile, target and the incident energy\cite{watanabe95}.
In this way, it was verified that QMD gives an adjustment-free
description of
the pre-equilibrium
(nucleon-in, nucleon-out) reactions at intermediate energy region in a
 unified manner.

\subsection{Step-wise contributions}

In order to have a better understanding of the reason why the QMD can
reproduce
the measured data so well, we compare in Fig. 4 the separate
contributions to $^{58}$Ni(p,xp')
cross sections from the
1-, 2- and 3-step processes and the total of all steps
calculated by the QMD theory with experimental data\cite{fortsch}.
Shown also are two
arrows $\alpha$ and $\beta$ corresponding to the angles expected from
the 1-step quasifree scattering process without and with the
acceleration effect
by the mean field, i.e.,

\begin{equation}
\cos \alpha~=~\sqrt{\frac{E_{\rm out}}{E_{\rm in}}},~~~
\cos \beta~=~\sqrt{\frac{E_{\rm out}~-~V}{E_{\rm in}~-~V}}
\label{eq:qfs}
\end{equation}
where $E_{\rm out}$ and $E_{\rm in}$ denote the energy of the outgoing
 and incoming
particles in the laboratory frame, respectively, while $V$ indicates
the
depth of the mean-field potential which has been taken to be $-50$
MeV.

Fig. 4 indicates the followings:
\begin{itemize}
\item the 1-step process is dominant at the forward angles, while
at badkward angles the 2- and 3-steps are responsible to reproduce the
 measured
cross sections.
\item the 1-step cross section does not have a peak neither at the
angle $\alpha$
nor $\beta$, instead it seems to have peaks at further forward angles
for
every secondary proton energy.  As will be shown later, it is the
Fermi
motion of the target nucleon that is responsible for the shift of
quasifree
peak toward the forward angles.
\item the 1-step cross section does not fall off at the very forward
angles
for a high energy ejectile, i.e., at $E_{\rm out}$ close to $E_{\rm
in}$.
This is a special feature of QMD theory, because the 1-step cross
section calculated
by the simple kinematical theory, as represented by the Kikuchi-Kawai
formula\cite{kikuchi}, drops off sharply at the forward angles, which
is the
reason why the cascade model often underpredicts the cross sections at
 this
angular region.  We will show later that the refraction of the
projectile
and the ejectile is responsible for not having the steep drop at the
forward angles.
\item the 1-step cross section has non-negligible contribution beyond
90-degree.
The momentum distribution, especially the high-momentum component, is
the
reason of this spreading out the quasifree peak toward the backward
angles.
\end{itemize}
Therefore, three effects are found to be important to reproduce
the measured (p,xp') cross section at the entire angular range; the
refraction,
the momentum distribution including the high-momentum component, and
the multi-step
contributions.  Effects of the refraction and the momentum
distribution will be
discussed further in later subsections.

\subsection{Comparison with Feshbach-Kerman-Koonin model predictions}

The step-wise contributions to $^{90}$Zr(p,xp') and (p,xn) reactions
for incident energy at 160 MeV, and
$^{27}$Al(p,xp') and (p,xn) reactions at 90 MeV calculated by the QMD
are compared in
Figs. 5, 6, 7 and 8 with those calculated by the multi-step direct FKK
 theory and
the experimental
data\cite{richter,scobel,wu,kalend}.  It is confirmed that both
theories can
reproduce the measured values rather satisfactorily.  The
similarity between the QMD
and the FKK results, as well as their abilities to reproduce the data,
are rather striking considering that these two theories are based on
completely
different concepts; the QMD being a superposition of N-N scattering
with
mean-field
effects, while the FKK is based on the DWBA scattering amplitudes.
A noticeable difference, however, exists at the lowest ejectile
energy of $^{90}$Zr(p,xp') and (p,xn) reactions, where the FKK
predictions are
bigger than the measured data at forward angles ($\theta \ge$
50-deg.), and
are smaller at backward angles.  The QMD results do not show such a
problem.
The main difference between the QMD and FKK calculations come from the
difference in the 1-step cross sections; the 1-step FKK cross section
has
a prominent peak at around 30-degrees, and drops off steeply at
backward
angles, while the 1-step QMD cross section has much flatter shape.
We will show later that the difference in the momentum distribution
has
little effect on the angular distribution shape from the 1-step
process.
Therefore we conclude that the difference in the 1-step QMD and FKK
cross sections
come from the difference in the angular distribution of the elementary
 process;
in the QMD calculation, the 1-step cross section is determined by the
N-N
cross section which is nearly isotropic in the CM of two colliding
nucleons,
while in the FKK theory it is
determined by the DWBA.  In spite of the difference in the 1-step
cross
sections, however,
the 2- and 3-step QMD and FKK angular distributions are very similar.
 This
will be another confirmation of the result obtained by Chadwick and
Oblo\v{z}insk\'y \cite{chadob} who have shown that the linear-momentum
dependent state
density obtained by the exact and statistical
Gaussian solutions become identital at 2p-2h and 3p-3h states in spite
 of
a difference in the 1p-1h state.

\subsection{Quasi-free scattering}

As shown in the previous sections, the 1-step quasifree scattering
(QFS) cross sections
calculated by the QMD theory has two prominent features; it does not
fall off at the
very forward angles unlike
the kinematical calculations\cite{kikuchi}, and it does not fall off
at the backward angles as rapidly as one predicted by the FKK theory.
 Here, we
investigate two items that may play important roles in
the quasifree scattering process;
the momentum distribution and the surface refraction effect.

First, we have investigated the effect of the momentum distribution to
QFS angular
distribution.  Figure 1 indicates that the momentum distribution in
the QMD calculation
differs noticeably from that of the uniform Fermi gas (UFG) model,
which was adopted
in Kikuchi-Kawai theory.  Instead of the
square-shaped
distribution, the momentum distribution in the QMD has a Gaussian-like
shape with
small portion above the Fermi momentum, which is in between the UFG
and the
2-Gauss distribution suggested by Haneishi and Fujita\cite{haneishi}.

We have compared the QMD angular distribution from the 1-step (p,n)
process of
$^{90}$Zr for incident energy at 160 MeV in the topmost parts of
Fig.~9 with a simple 1-step Monte-Carlo
calculations
with momentum distributions of UFG and 2-Gauss.
The 1-step calculation was performed as follows:

\begin{enumerate}
\item Select energy of a neutron in the target either from
the UFG or from the 2-Gauss distributions, assuming a
nucleon binding energy of 8MeV and Fermi energy ($E_{\rm F}$) of
40MeV.

\item Make an isotropic scattering in the CM system of the
projectile and the selected neutron in the target.

\item The Pauli blocking effect is taken into consideration with a
blocking probability given by
\begin{equation}
P_{\rm block}~=~1-\left[ 1-\theta(E_{\rm F}-E_{1}') \right] \cdot
\left[ 1-\theta(E_{\rm F}-E_{2}') \right]
\label{eq:pauli}
\end{equation}
where $E_{1}'$ and $E_{2}'$ denote the energies of scattered
particles.

\item If the collision is not blocked, the energy and angle of the
scattered particle (which originally was in the target) in the
laboratory frame is recorded.

\item Repeat items 1 to 4 many times.

\item The absolute magnitudes of these 1-step cross section was
normalized
to the corresponding 1-step QMD cross section.

\end{enumerate}

The upper two figures in Fig. 9 show that, in the main part of the
angular distribution,
the difference between the UFG and the 2-Gauss results is not very
noticeable.  The main
difference lies at the very forward and backward angles, where UFG
result exhibits
a steep drop, while the 2-Gauss result shows a slower decrease.  This
is definitely
due to the high-momentum component in the 2-Gauss momentum
distribution, because
this difference disappears when
we cut the high-momentum component in the 2-Gauss distribution .
Also, it is clear that the two distributions do not give the QFS peak
at angles
denoted by $\alpha$ nor $\beta$, but give a peak at
more forward angles.  Therefore,
the Fermi motion of target nucleons was found to shift the QFS peak to
 the forward
angles.  The peaks are, however, not very prominent in both cases;
the Fermi
motion tends to wash-out the QFS peak.  The 1-step QMD cross section
is in very
good accord with both 1-step results at intermediate angles.
As a matter
of fact, the angle beyond which the QMD cross section vanishes lies in
 between
the corresponding angles of the UFG and 2-Gauss results, because the
momentum
distribution in QMD lies in between these two distributions as shown
in Fig.~1.
However, the QMD results do not show decreasing angular shapes toward
0-degree.  Therefore, the reason why the 1-step QMD results
have large
cross sections in the vicinity of 0-degree was not explained by the
Fermi
motion of the target nucleons.

In the lower parts of Fig. 9, we have compared two kinds of 1-step QMD
cross sections, one with the full calculation, and one which cuts the
refraction
effects.  The 1-step QMD results without the refraction show a shape
very similar to
the one calculated with the uniform Fermi gas model, with a steep drop
 at the
very forward angles, while the refraction effect totally washes out
this steep
decrease.  Thus, it became clear now that it
is the refraction effect by the
mean-field which causes a non-decreasing 1-step cross sections in the
QMD
calculation at 0-degree region.   This effect, together with
contributions from the 2-, 3- and higher steps,
makes the total QMD cross sections to have a smoothly varying angular
shape
from the very forward to backward angles, which is in good accord with
 the
measured data.  Therefore, we conclude that the Fermi motion of target
nucleons, mean
field refraction,
and the multi-step effects are essential in predicting the angular
distributions of
preequilibrium (N,N') cross section in this energy range.

In the present calculation, the mean-field refraction effect
washes out the decrease of the cross section at the very
forward angle.  This result, at first glance, may look completely
opposite to the one obtained with the Geometry-Dependent Hybrid
model (GDH) where the steep increase at the forward angle is
washed out by the surface refraction at low incident
energy (e.g., Fig. 4 of Ref. \cite{blann84}).
At higher incident energy, however, the GDH predicts decreasing
angular shape toward 0-degree that is smeared out when the
refraction effect is considered (20 MeV n of Fig. 6 in Ref.
\cite{blann84}), which is in good agreement with the present
result.
The reason
why the GDH calculation has such a steep increase at the
forward angle of low-incident energy is unclear to us: at
least the 1-step process
is concerned, the cross section must have a decrease as
a result of kinematical restriction and Pauli-principle
as Kikuchi-Kawai formula indicates.  It may be just a
result of finite angle-binning carried out by Blann et al.
In the QMD calculation, both
the Fermi motion and the surface refraction effect
are taken into account by means of the equation-of-motion
(i.e., Eq. \ref{eq:eos}) in a unified manner, and any
arbitrariness is not involved as is the case of GDH
to introduce the
surface refraction.

It must be
noticed that the refraction effect in our calculation might be
overemphasized
due to the fact that momentum dependence is not included in the
effective
N-N force that changes the mean-field potential
from attractive at low energy to repulsive at energies higher
than approximately 200 to 300 MeV region.
However,
the relativistic approach on the optical potential
gives a wine-bottle-bottom shaped potential that remains attractive
at the surface regime even at high energy region where the
potential at the nuclear interior becomes significantly
repulsive\cite{rikus,hama}.
Moreover, according to Gadioli
and Hodgson\cite{gadioli}, inclusion of the momentum-dependent
potential
leads to two opposite consequences: (a) it reduces the importance of
refractions
of nucleons by reducing the potential at 200-300MeV region.  This
effect increases
the probability of emission of particles.   However, (b) the particles
 in the
mean-field will have on the average lower kinetic energies, and (due
to the
increase of the N-N cross section) smaller mean free paths with the
consequence
that the nuclear transparency is decreased.  These two effects tends
to cancel
 each
other, and the net effect of the momentum-dependent potential
might be substantially reduced.

\subsection{Multiple preequilibrium emission}

In Ref. \cite{chadchiba}, we have shown that the QMD gives results
consistent with the FKK theory for the energy spectra of first and
second
particles
emitted from the preequilibrium process up to projectile energy of 200
 MeV.
Recently, there is a growing interest
in nucleon-induced reactions up to 1 GeV region stimulated by the
results obtained at LAMPF facility on the Pb(n,x) and Fe(p,x)
reactions\cite{vonach94,vonach95} and from a practical point of
view\cite{nea,nagel}.
Our QMD approach can be applied to above 1 GeV without any change,
and there is no
limitation in the number of particles emitted from the preequilibrium
process.  Therefore we have extended the analysis given in Ref.
\cite{chadchiba}
up to 1 GeV, and investigated the importance of MPE process as a
function of
projectile energy.

We have shown in Fig.~10 the QMD results for the fractional
contributions to the total number of
inclusive proton
emission contributed by the particles emitted as the 1st, 2nd, 3rd,
4th
and 5th particles.  The upper figure shows the percentage
when all emission energies
are considered, while the lower figure includes only emissions of
above
25 MeV (as defined in the preequilibrium regime in Ref.
\cite{chadchiba}).
The lower figure shows that sum of the contribution by the 3rd, 4th
and 5th particles
in the preequilibrium process occupies a fraction of about 30 \% at
the incident
energy of 500MeV, and almost 50 \% at 1 GeV, showing a clear necessity
 of
including the MPE process more than 2 particles.

\section{Summary}

We have shown that the Quantum Molecular Dynamics (QMD) model can
reproduce the measured data for intermediate-energy
nucleon-induced preequilibrium nucleon-emission process without
any adjustment of the underlying parameters.  Based on this success,
we then
have studied some of the open problems in the preequilbrium reactions;
the angular distribution and the multiple preequilibrium particle
emission (MPE).
The QMD calculation has not shown the prominent quasifree scatterint
peak, which is consistent with the measured data.  The reason of
the overall agreement with the data was explained by the
Fermi motion of target nucleons, the refraction of projectile and
ejectile, and
contribution from the multi-step processes.  The MPE process beyond
2-particle emission
was found to exceed 30 \% at 500MeV and reaches almost 50 \% at 1 GeV,
 thus
becoming the major reaction mechanism at this energy region.

\acknowledgements

The authors would like to thank Prof. M. Kawai and Dr. Y. Watanabe
of Kyushu University, Prof. H.
Horiuchi and Dr. E. Tanaka of Kyoto University
for valuable comments
and discussions.


\begin{figure}
\caption{Nucleon momentum distribution of $^{90}$Zr.  The solid
histogram presents the
resutls of QMD calculation.  The smooth broken curve and the square
distributions
designates the 2-Gauss distribution parametrized by Haneishi and
Fujita [15]
and uniform Fermi gas distribution, respectively.}
\end{figure}

\begin{figure}
\caption{The $^{58}$Ni(p,xp') cross sections at $E_p$ = 120 MeV (left)
 and 200 MeV (right).
The data have
been multiplied by the amount denoted in the parentheses.}
\end{figure}

\begin{figure}
\caption{The $^{90}$Zr(p,xp') and (p,xn) cross sections at $E_p$ = 160
 MeV.  The data have
been multiplied
by the amount denoted in the parentheses.}
\end{figure}

\begin{figure}
\caption{The $^{58}$Ni(p,xp') cross sections at $E_p$ = 120 and 200
MeV.  The
total (thick solid line), 1-step (dashed line), 2-step (broken line)
and 3-step
(thin solid line)
QMD cross sections are compared with experimental data.  The arrows
$\alpha$ and $\beta$
denote the position of QFS peaks as given by Eq. (9).}
\end{figure}

\begin{figure}
\caption{The $^{90}$Zr(p,xp') cross sections at $E_p$ = 160 MeV.  The
left row compares
the predictions of
total (thick solid line), 1-step (dashed line), 2-step (broken line)
and 3-step
(thin solid line)
QMD cross sections with experimental data, while the right one those
of FKK theory with
experimental data.}
\end{figure}

\begin{figure}
\caption{The $^{90}$Zr(p,xn) cross sections at $E_p$ = 160 MeV.  The
left row compares the predictions of
total (thick solid line), 1-step (dashed line), 2-step (broken line)
and 3-step (thin solid line)
QMD cross sections with experimental data, while the right one those
of FKK theory with
experimental data.}
\end{figure}

\begin{figure}
\caption{The $^{27}$Al(p,xp') cross sections at $E_p$ = 90 MeV.  The
left row compares the predictions of
total (thick solid line), 1-step (dashed line), 2-step (broken line)
and 3-step (thin solid line)
QMD cross sections with experimental data, while the right one those
of FKK theory with
experimental data.}
\end{figure}

\begin{figure}
\caption{The $^{27}$Al(p,xn) cross sections at $E_p$ = 90 MeV.  The
left row compares the predictions of
total (thick solid line), 1-step (dashed line), 2-step (broken line)
and 3-step (thin solid line)
QMD cross sections with experimental data, while the right one those
of FKK theory with
experimental data.}
\end{figure}

\begin{figure}
\caption{The $^{90}$Zr(p,xn) cross sections at $E_{p}$ = 160 MeV and
$E_{n}$ = 120 (left
figures) and $E_{n}$ = 40 (right )MeV.  The upper two figures compare
 1-step QMD cross section
 (solid lines)
with a simple 1-step Monte-Carlo calculations assuming
the uniform Fermi gas
(broken histograms)
and 2-Gauss (long-broken histograms) momentum distributions of target
nucleons.  The lower two figures compare the full 1-step QMD
cross section with a calculation ignoring the refraction effects.}
\end{figure}

\begin{figure}
\caption{Relative contributions of multi-particle emissions as a
function of the
incident energy calculated by the QMD.  The upper figure includes all
emission energies,
whereas the lower figure considers only emission energies above 25 MeV
 (the
preequilibrium regime).}
\end{figure}

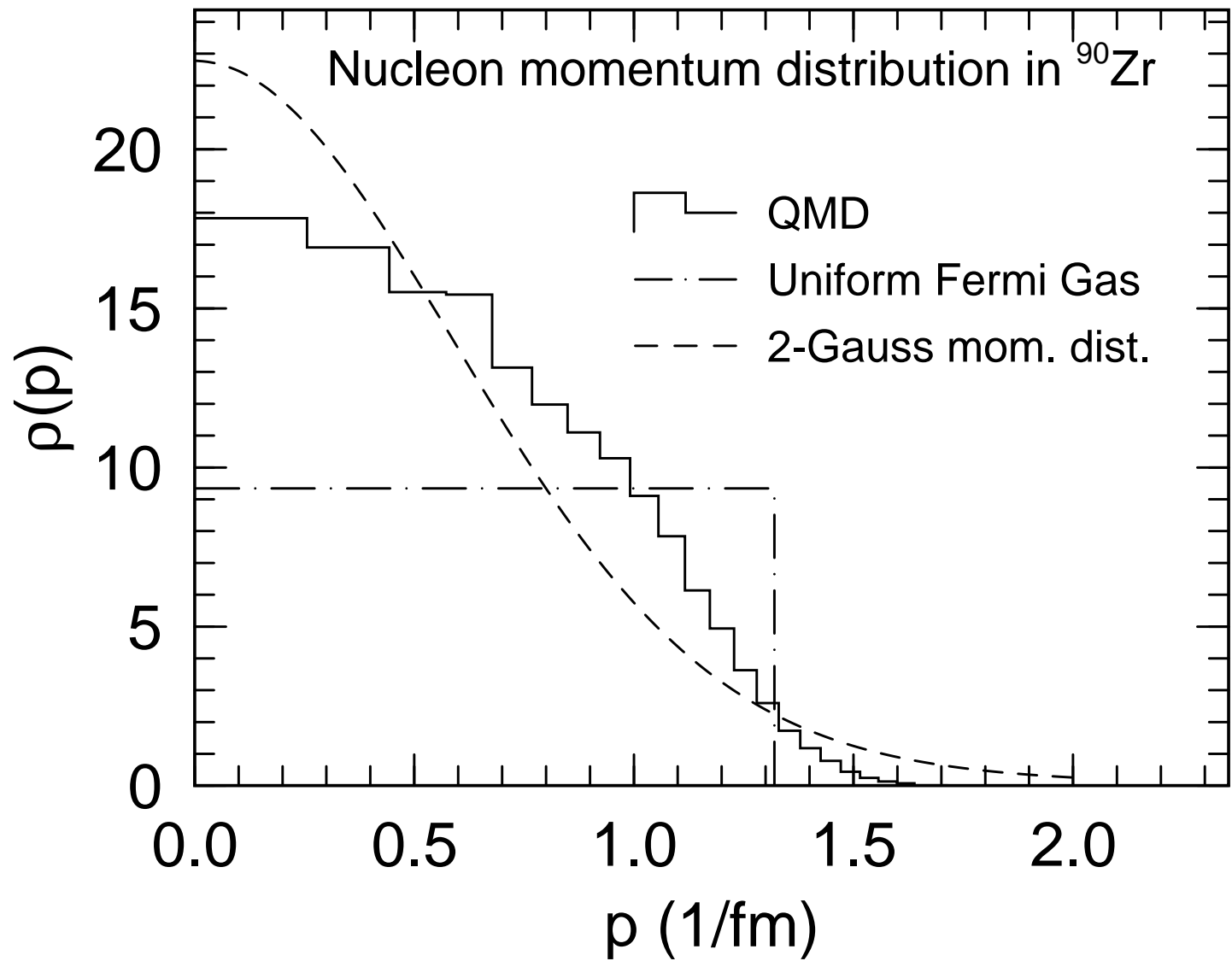
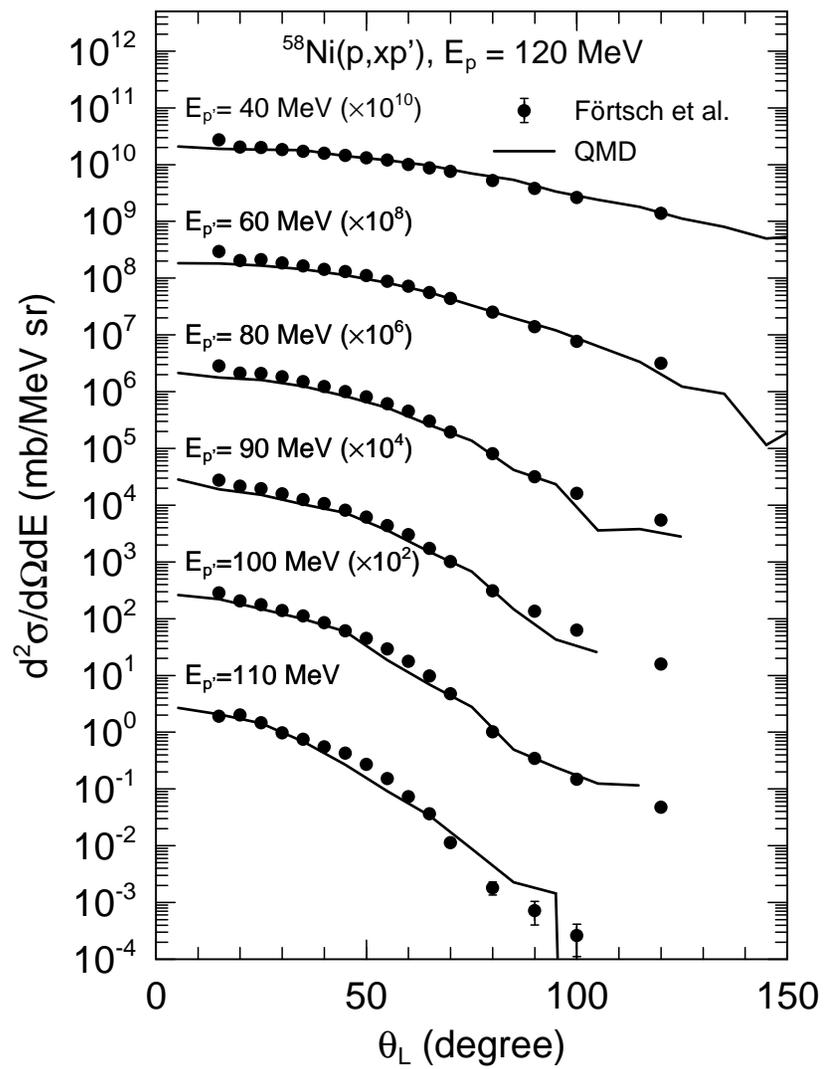
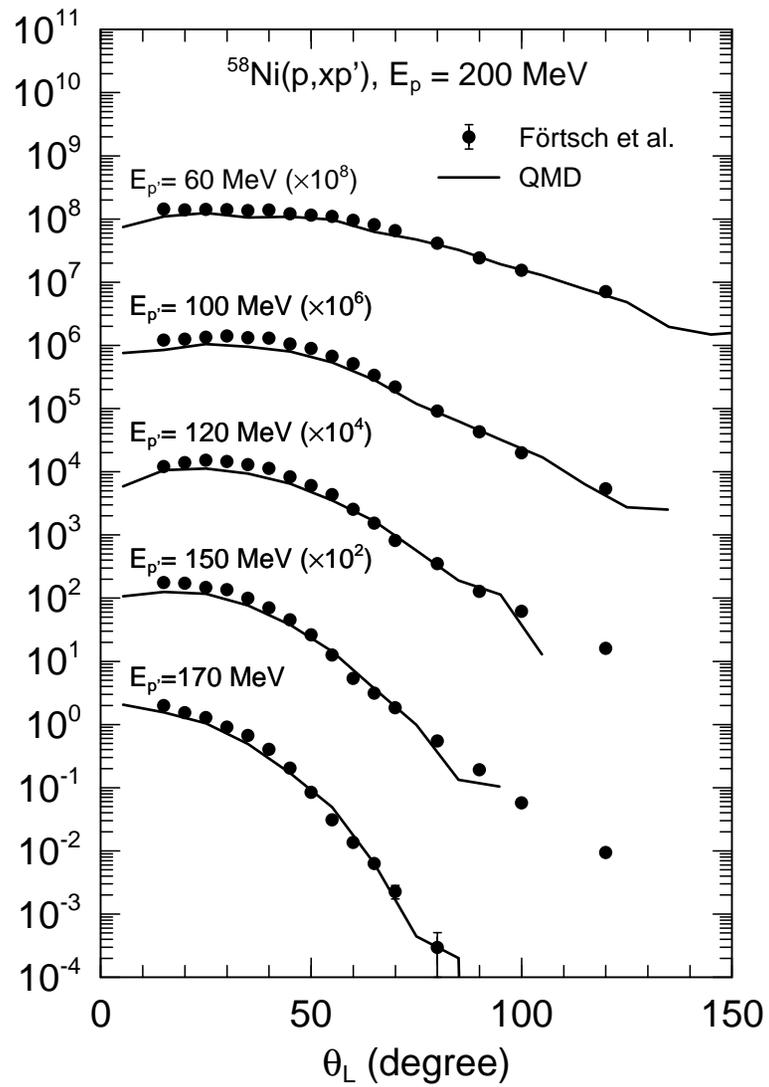
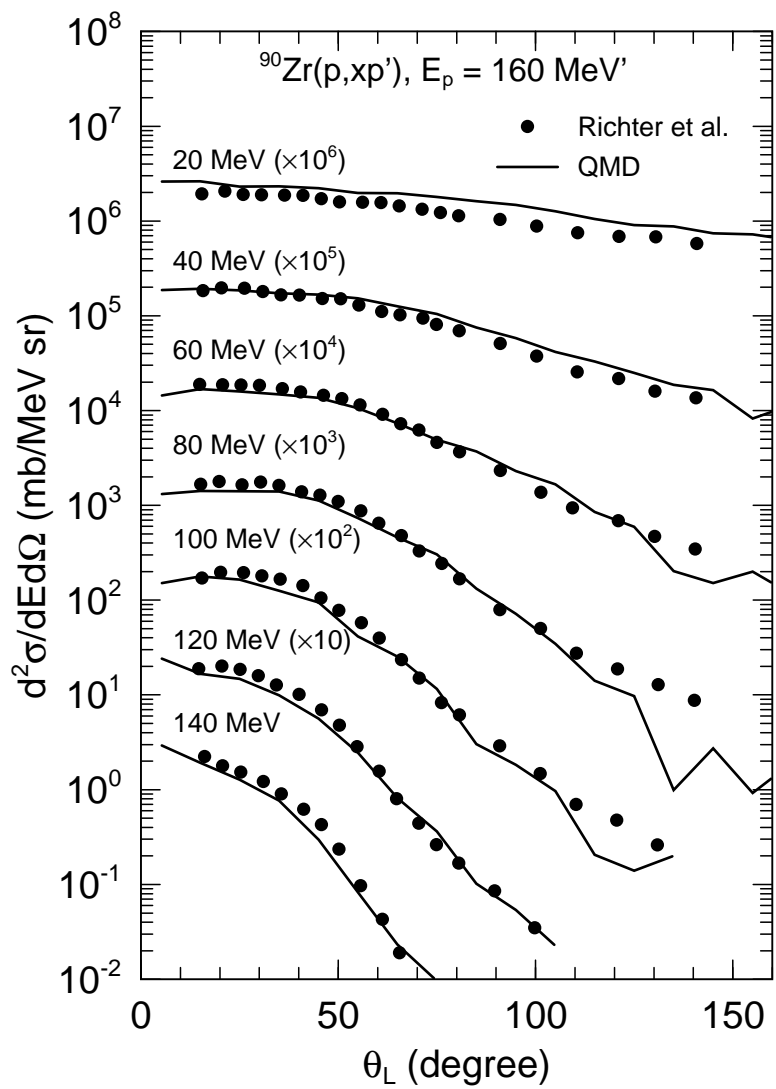
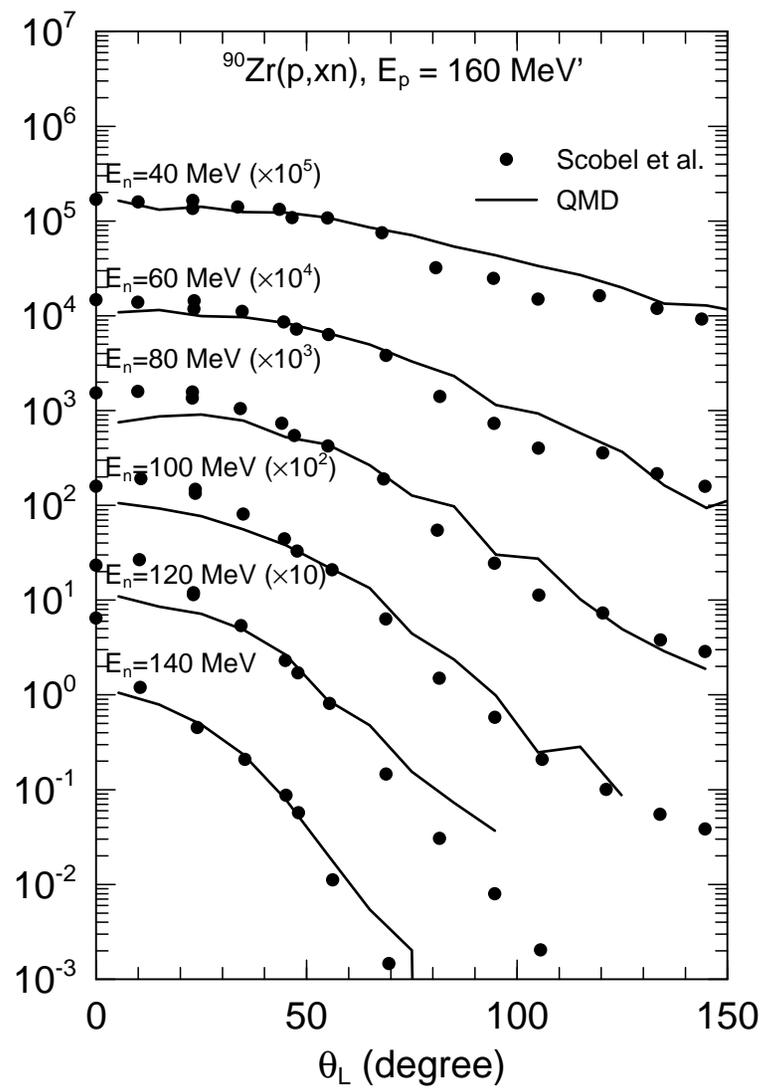
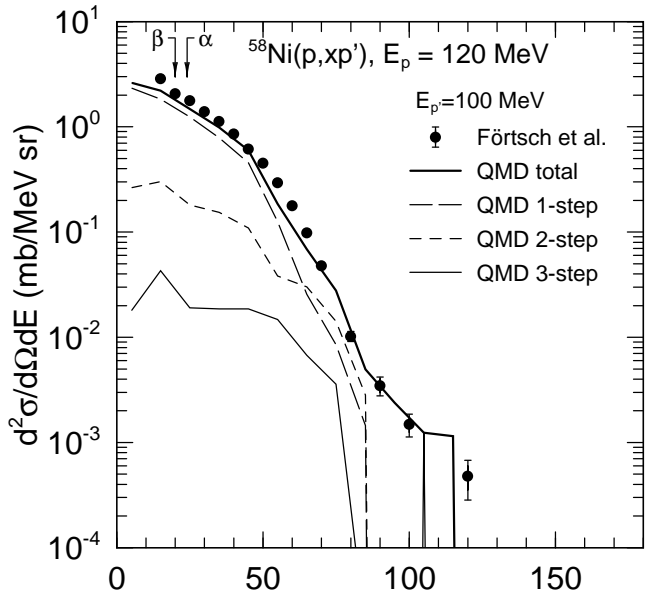
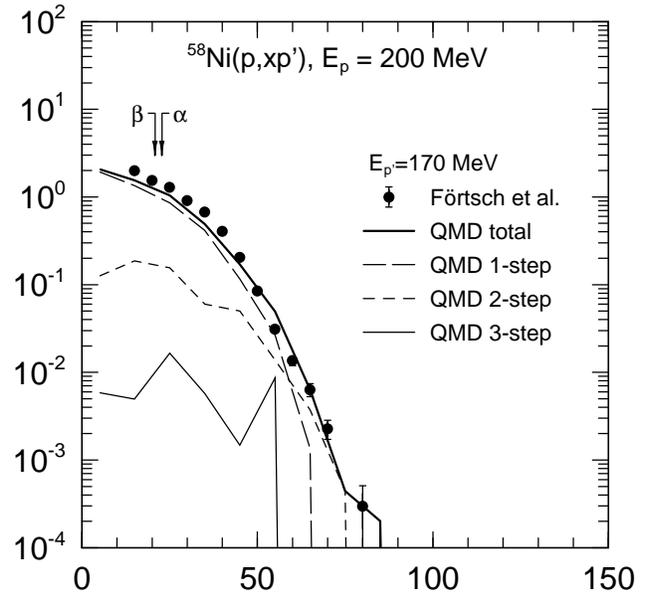
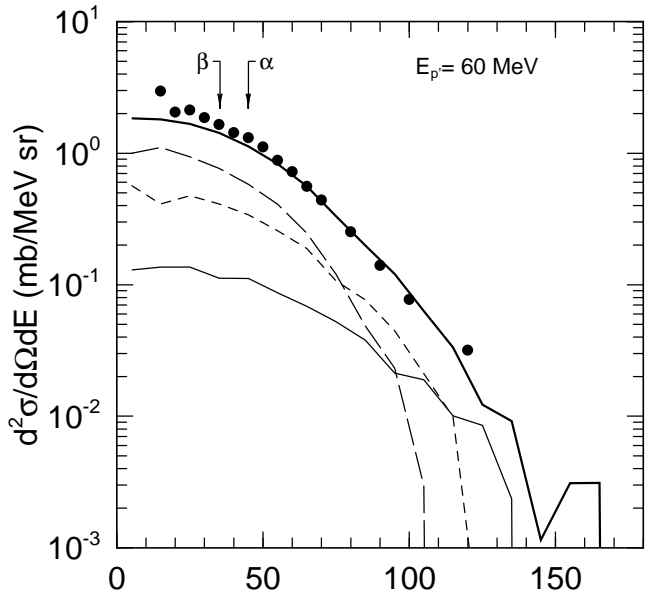
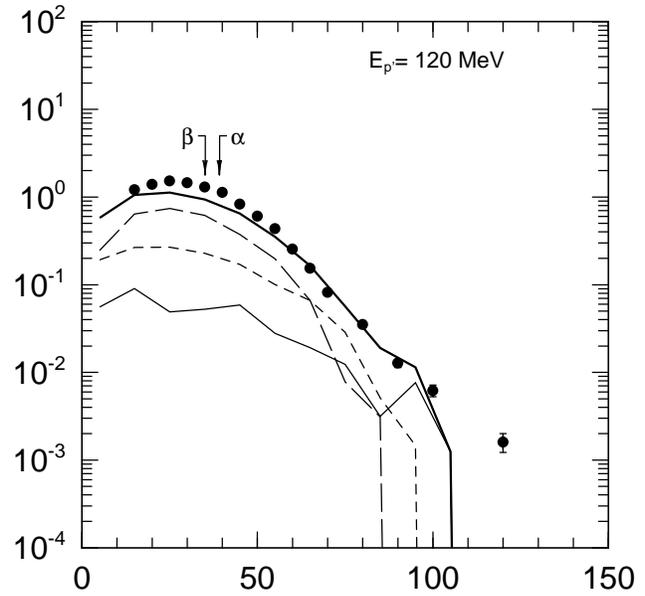
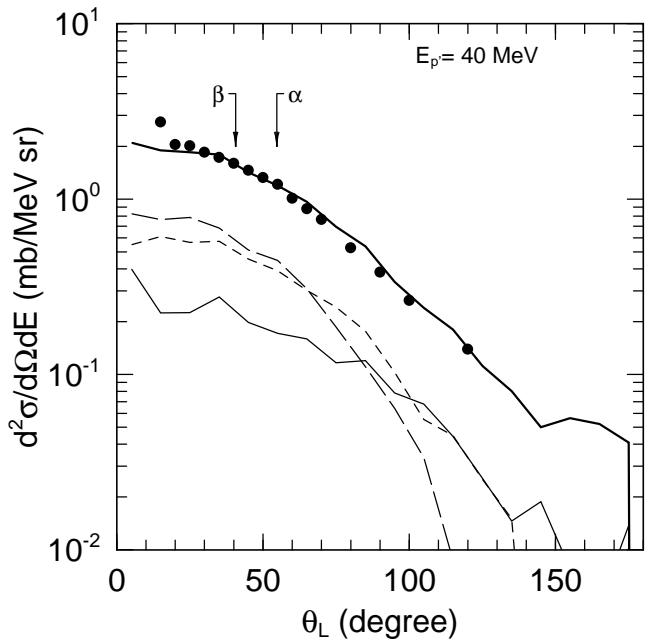
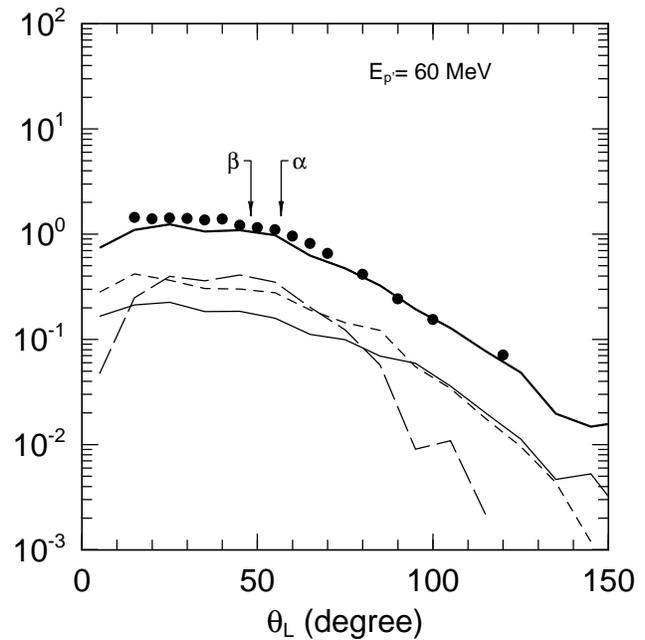
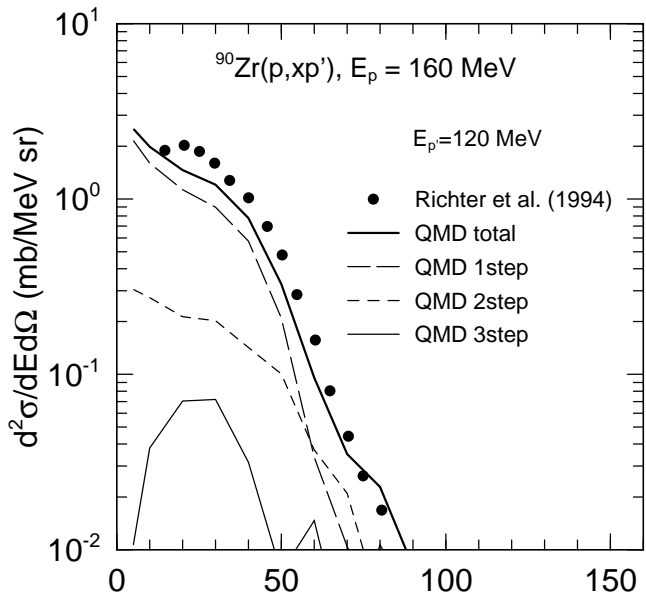
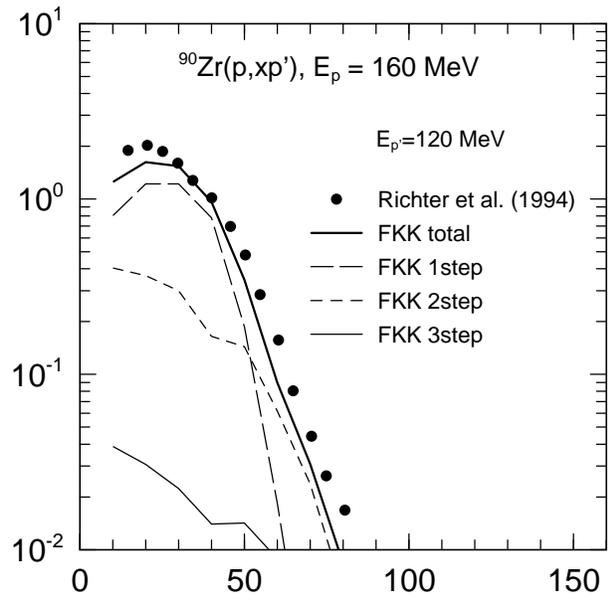
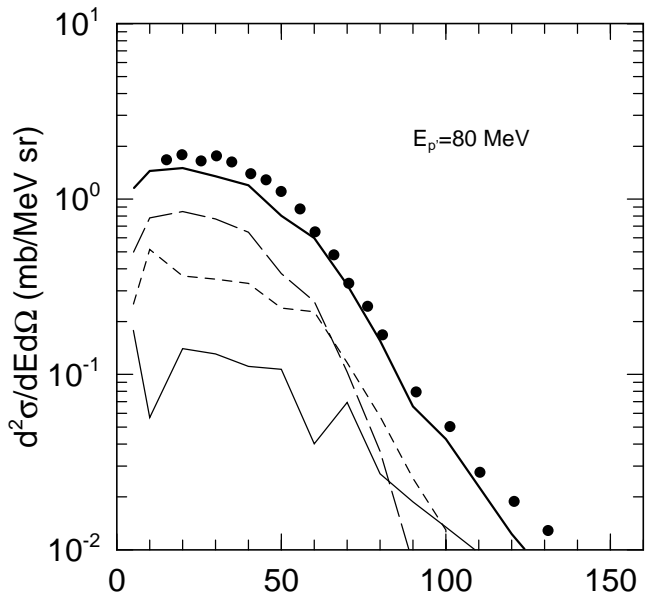
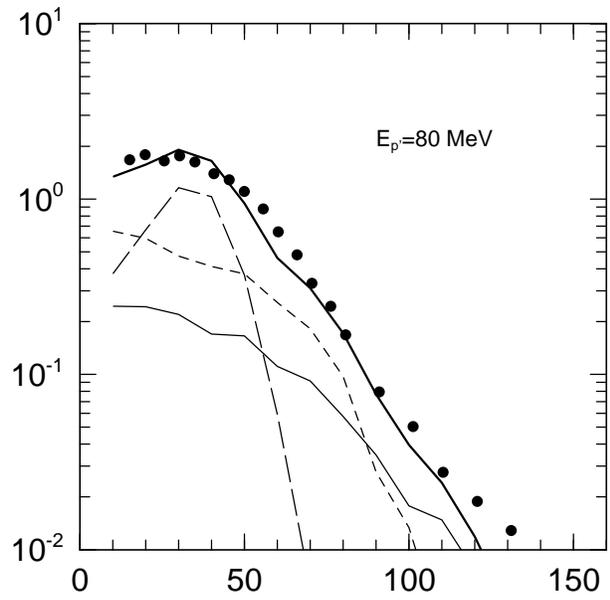
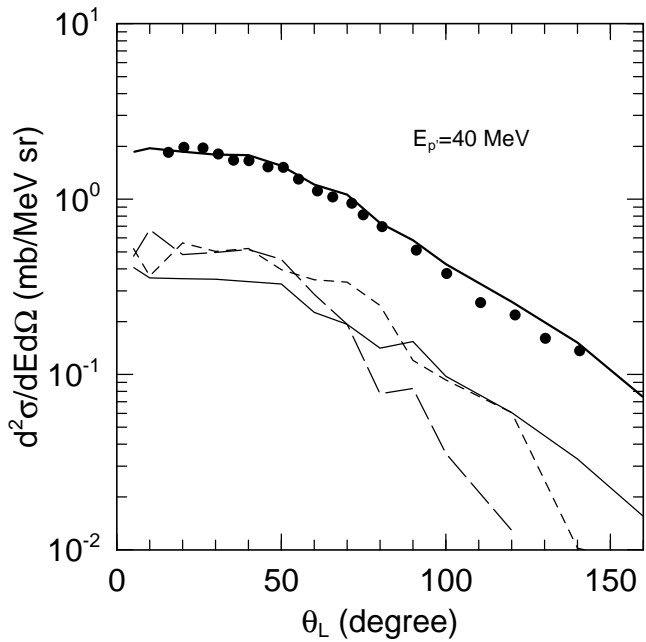
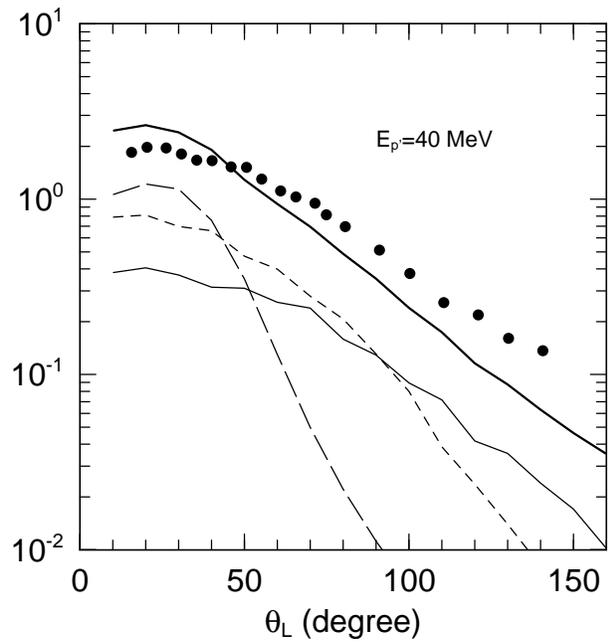
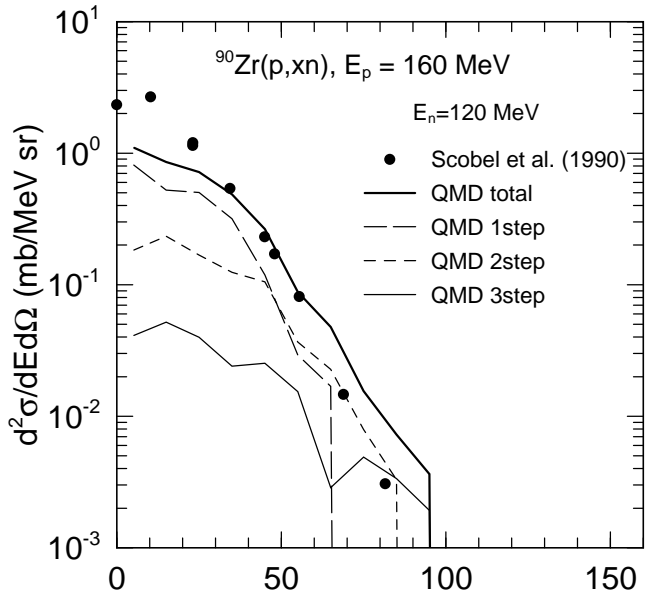
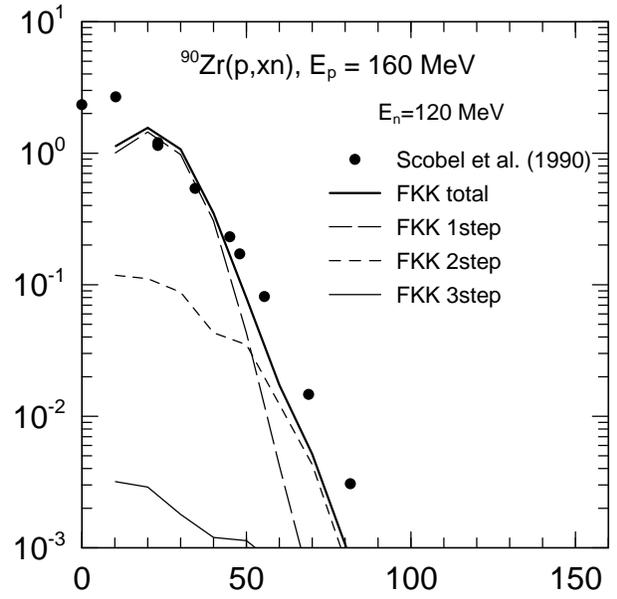
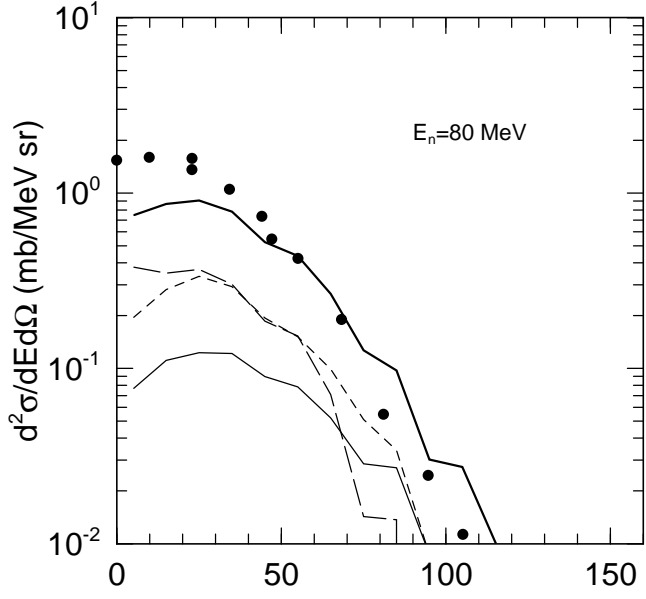
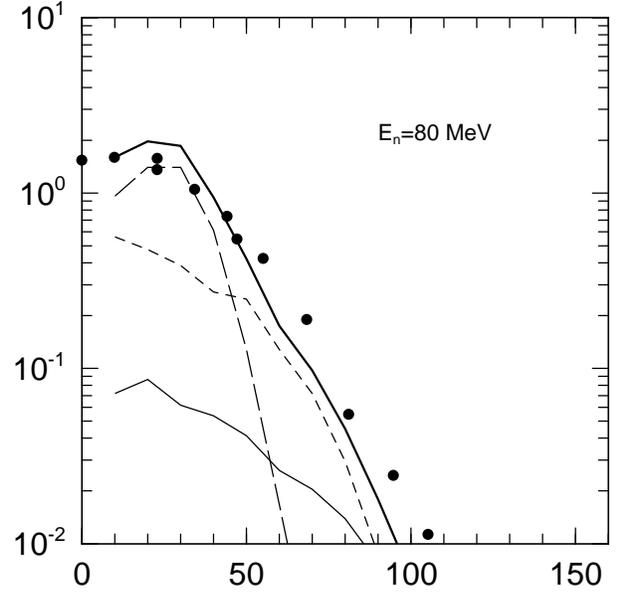
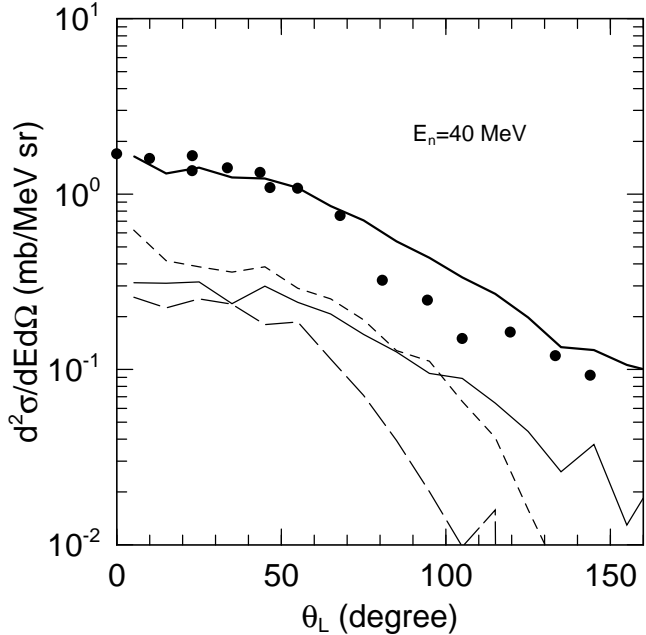
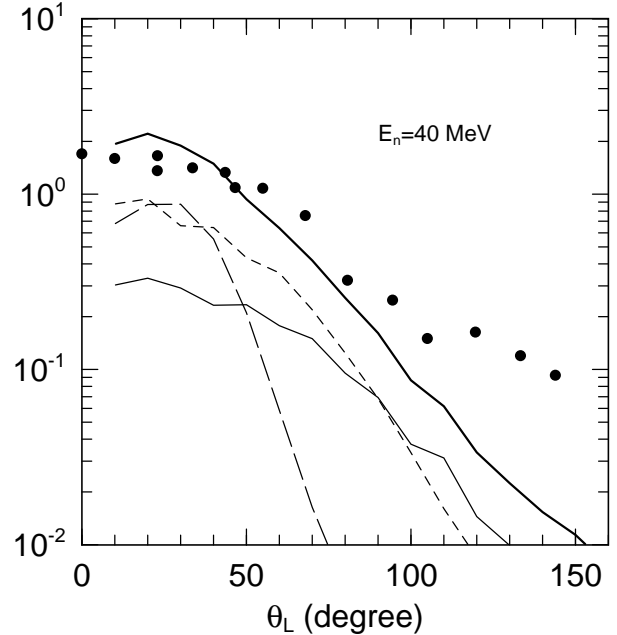
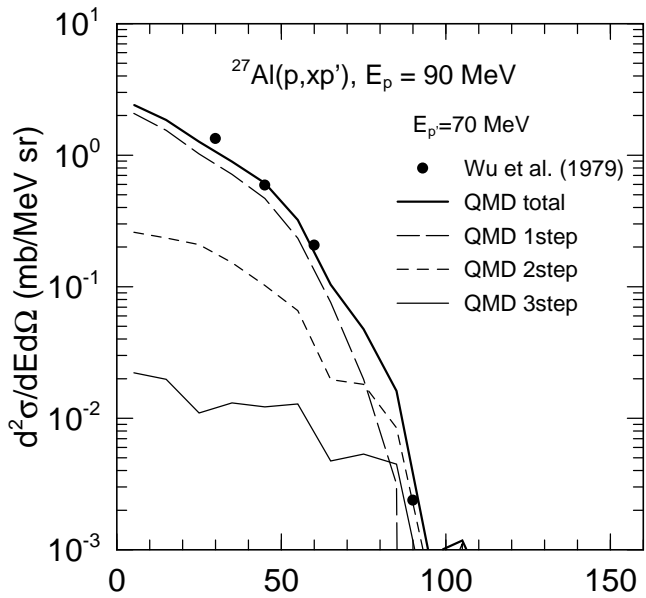
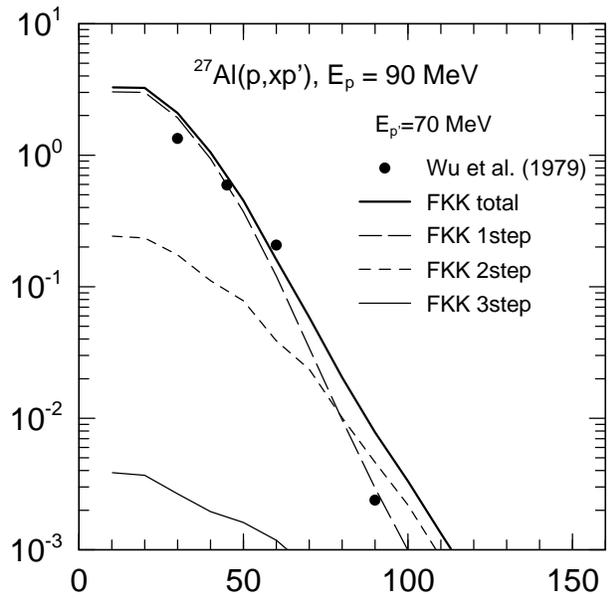
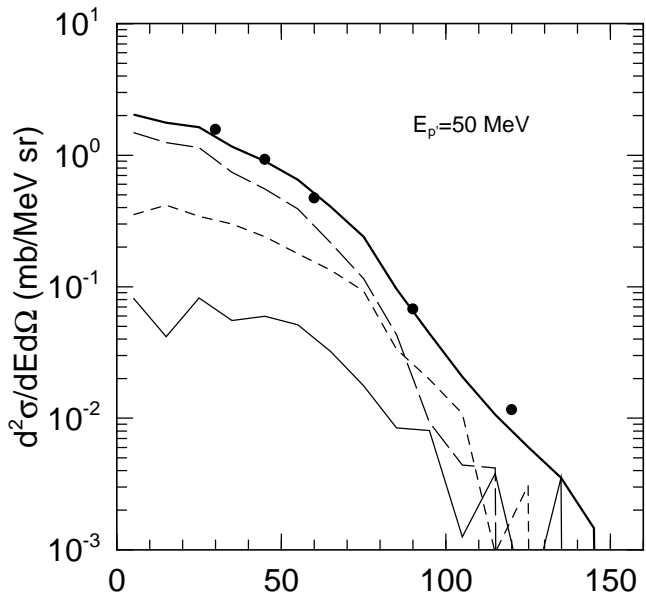
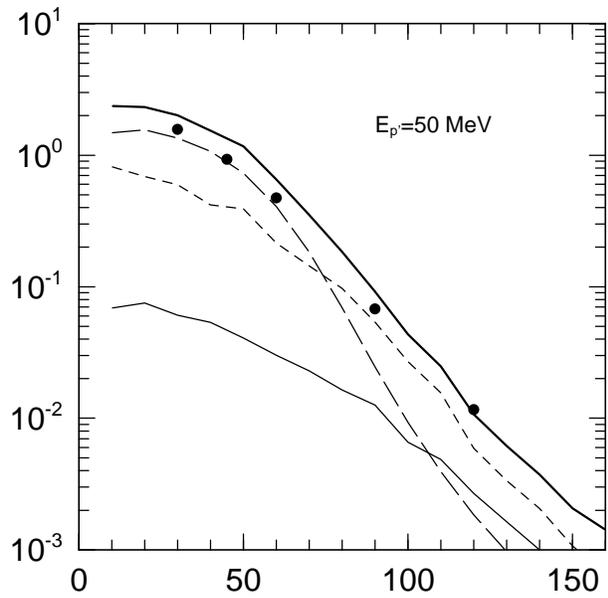
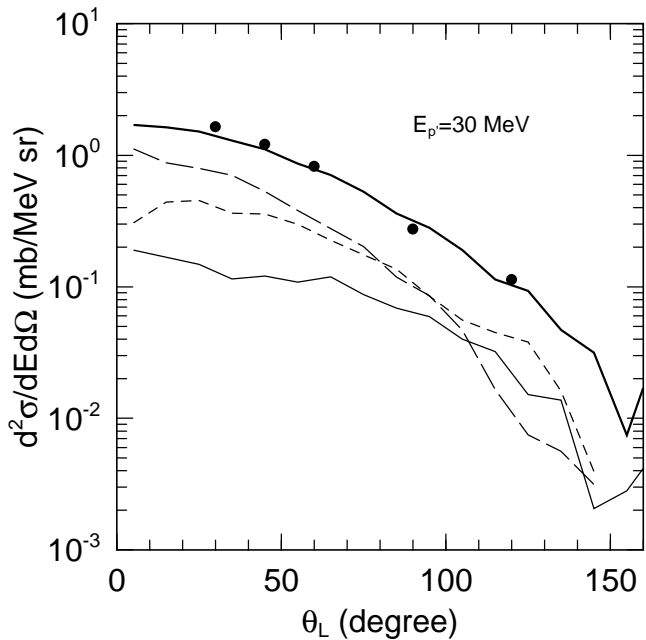
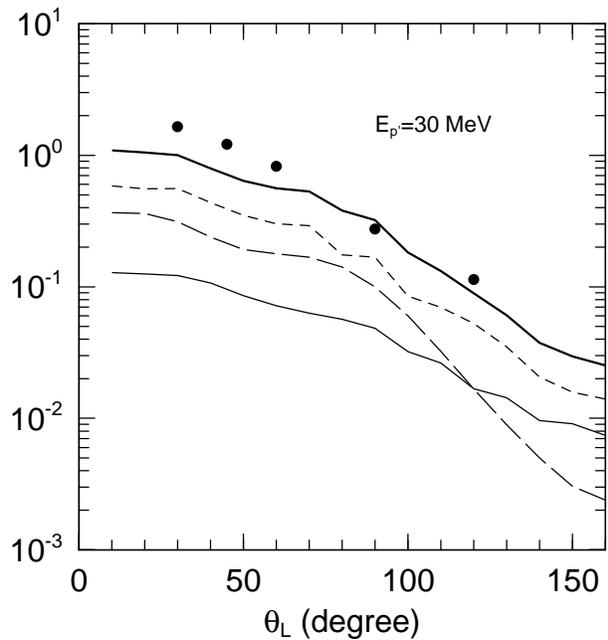
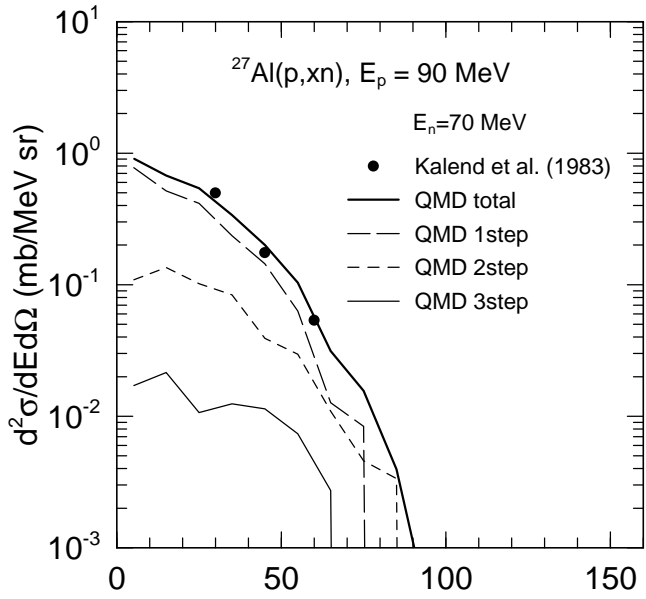
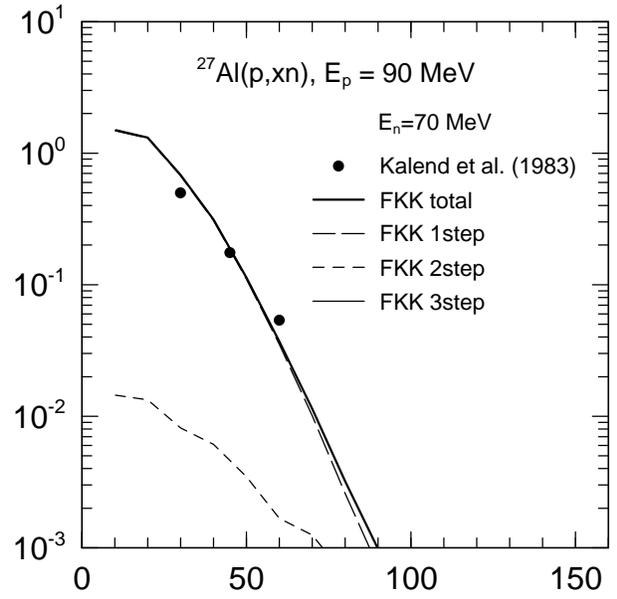
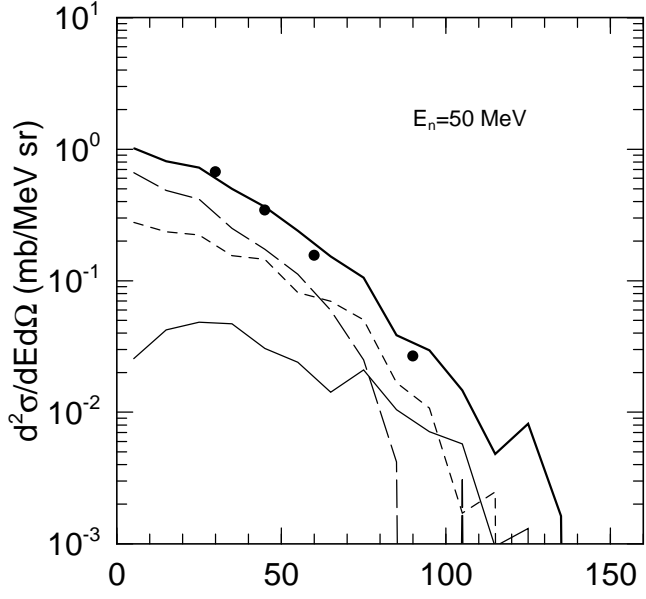
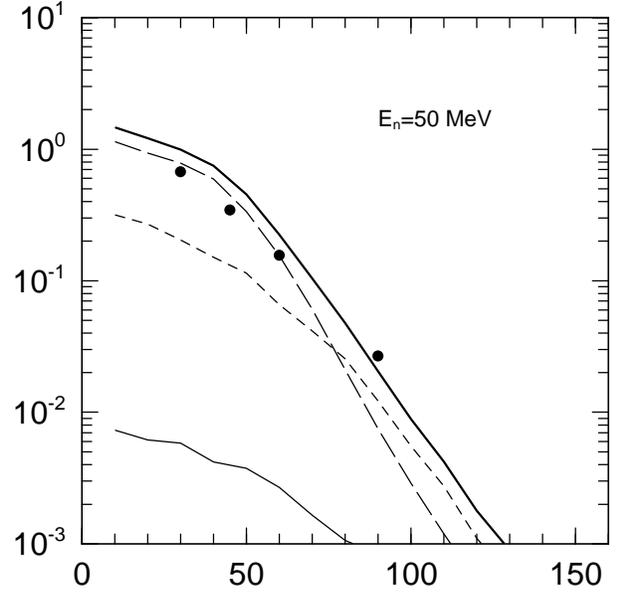
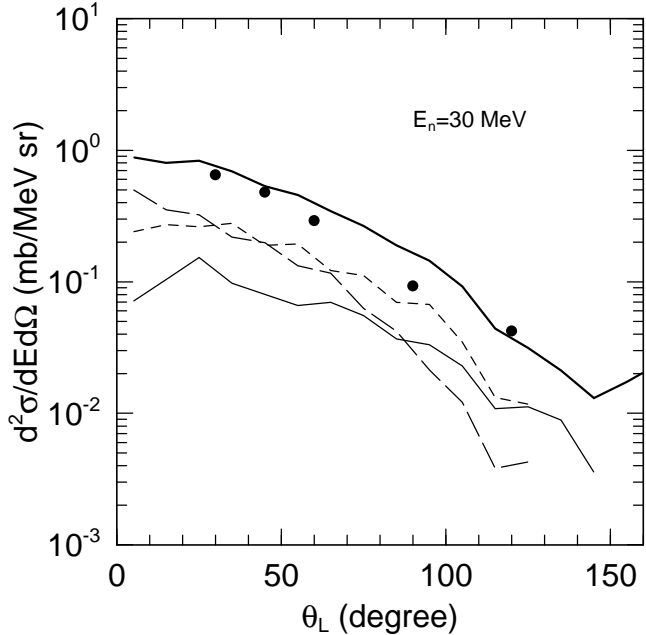
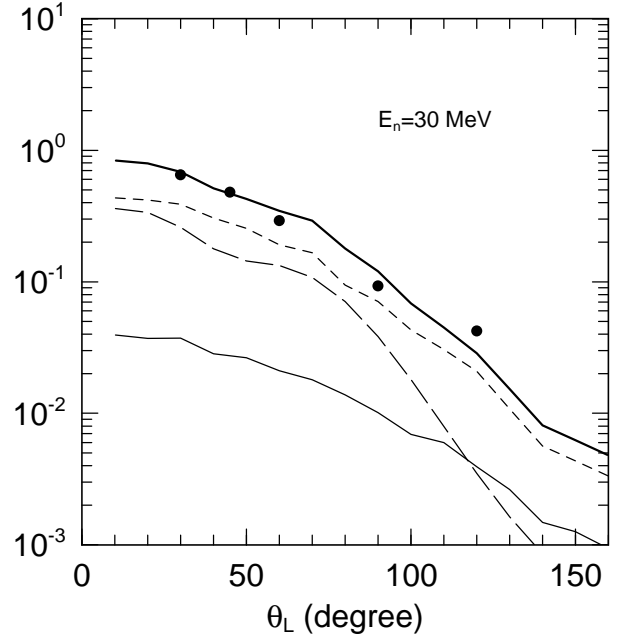
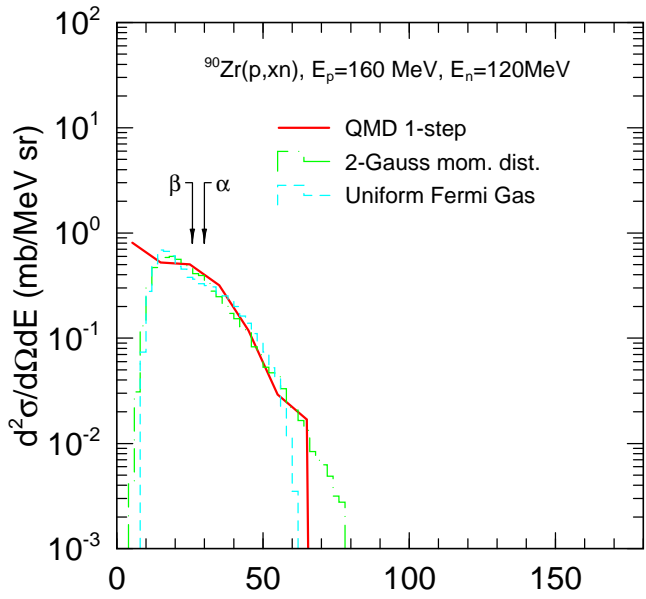
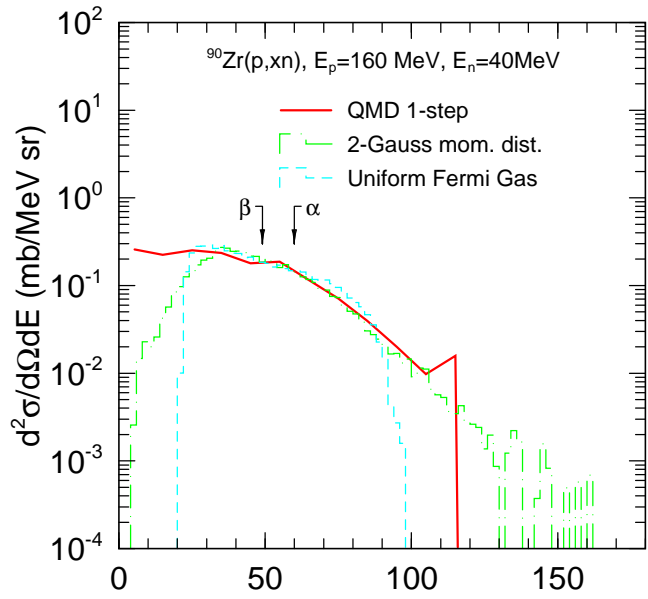
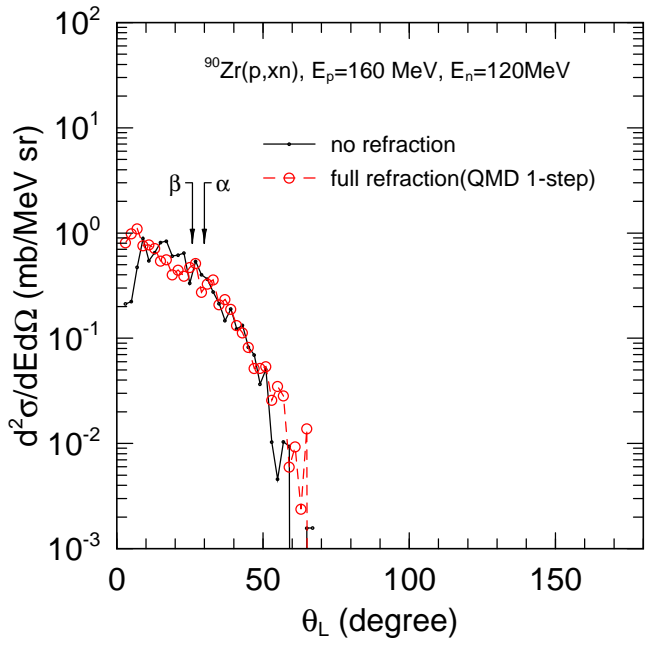
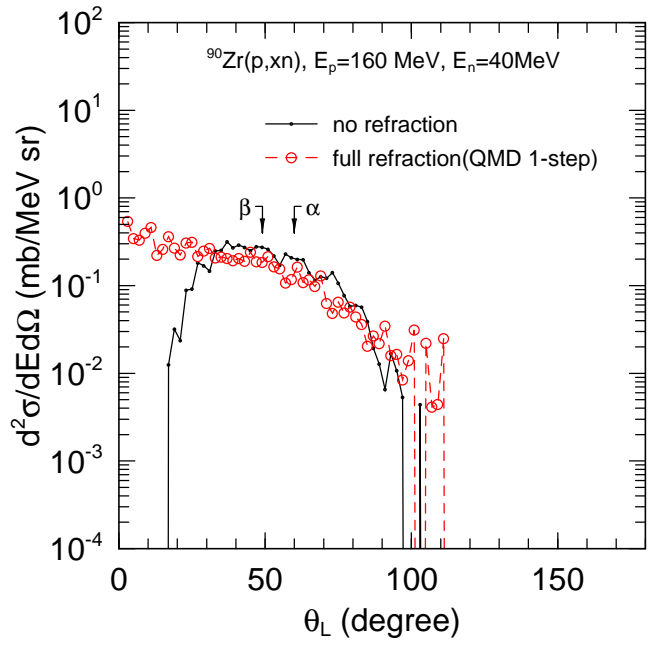
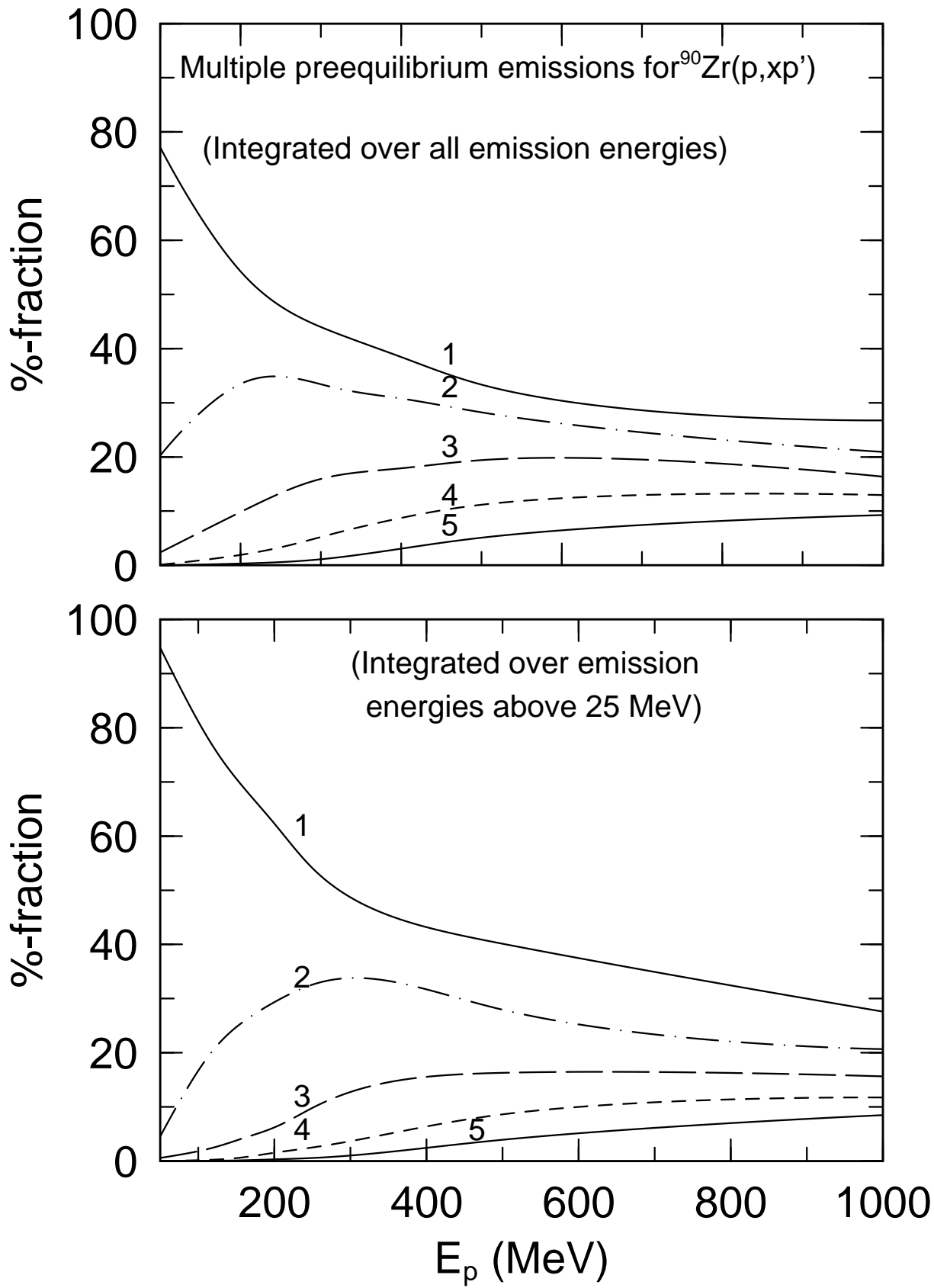
\end{document}